\documentclass[11pt]{article}
\usepackage{shuffle}
\usepackage{jheppub}
\usepackage{wasysym}
\usepackage{tabularx}
\usepackage{physics}
\usepackage{tikz}
\usepackage{graphicx} 
\usepackage{caption}
\usepackage{subcaption}

\usepackage{soul}

\usepackage{ae}

\usepackage{tabularx}

\usepackage{graphicx}
\graphicspath{{figures/}}
\usepackage{float}
\usepackage{graphbox}

\usepackage{amsfonts,amsmath,amsthm,amssymb,amsbsy}

\usepackage{mathtools}
\usetikzlibrary{calc,decorations.markings, shapes.misc}

\newcommand{\dlog}{\dd \hspace{-0.07cm}\log}

\title{Generalised Cluster Adjacency for Cosmology}
\author[]{Mattia Capuano,}\emailAdd{m.capuano@herts.ac.uk}
\author[]{Livia Ferro,}\emailAdd{l.ferro@herts.ac.uk}
\author[]{Tomasz \L ukowski,}\emailAdd{t.lukowski@herts.ac.uk}
\author[]{Alessandro Palazio}\emailAdd{a.palazio@herts.ac.uk}
\author[]{and Yao-Qi Zhang}\emailAdd{y.zhang59@herts.ac.uk}

\affiliation[]{Department of Physics, Astronomy and Mathematics, \\ University of Hertfordshire, \\  Hatfield, Hertfordshire, AL10 9AB, United Kingdom}

\abstract{In this paper we study the cluster algebraic properties of wavefunction coefficients for conformally coupled scalar theories in de Sitter cosmology. We show that the symbol of the wavefunction coefficient of the $n$-site path graph $P_n$ obeys a generalisation of cluster adjacency, where all letters in a given word belong to the same cluster of an $A_{2n-3}$ algebra, with certain additional constraints on the order of the letters. We call this property the ordered single cluster condition, and provide its physical interpretation. This condition is stronger than the usual cluster adjacency obeyed by neighbouring letters, and thereby constrains the symbol bootstrap far more tightly. We also show that for an arbitrary graph the alphabet carries a cluster-like structure, described by tubes and tubings on the graph, which allows for a similar bootstrap approach.

}

\begin{document}

\maketitle


\section{Introduction}

Cosmological correlation functions are central observables in modern cosmology, as they encode the physics of the early universe.
Traditional techniques for computing them quickly become intractable.  Recent developments have emerged by transferring mathematical methods from scattering amplitudes in maximally supersymmetric Yang-Mills theory (SYM) to the calculation of cosmological correlators in power-law Friedmann-Robertson-Walker (FRW) cosmologies.
In particular, progress has been made by considering the wavefunction coefficients, which are the building blocks of the wavefunction of the universe. These coefficients have been related to positive geometries, such as the cosmological polytope  \cite{Arkani-Hamed:2017fdk,Arkani-Hamed:2018bjr,Benincasa:2024leu}  and the cosmohedron \cite{Arkani-Hamed:2024jbp,Glew:2025otn,Ardila-Mantilla:2026cbo}; moreover, differential equations for these coefficients, derived using various methods, have been proposed  \cite{De:2023xue,De:2024zic,Arkani-Hamed:2023kig,Arkani-Hamed:2023bsv,Baumann:2024mvm,Baumann:2025qjx,Fan:2024iek,Grimm:2024mbw,He:2024olr,Benincasa:2024ptf,Hang:2024xas,Fevola:2024nzj,Capuano:2025ehm}.
Another powerful mathematical structure which appears for scattering amplitudes in SYM  \cite{Golden:2013xva}  is that of cluster algebras \cite{fomin2001clusteralgebrasifoundations,williams2013clusteralgebrasintroduction}. Their emergence plays a central role as they inform not only which singularities can appear, but also how these interplay.

The cluster algebraic structure of  cosmological correlators has remained hidden  until very recently. In  \cite{Capuano:2025myy,Mazloumi:2025pmx} it was shown that the singularities of the wavefunction associated with an $n$-site path graph $P_n$ are related to coordinates of the cluster algebra $A_{2n-2}$. 
However, for scattering amplitudes this is not the full story. Amplitudes exhibit cluster adjacency \cite{Drummond:2017ssj}, the statement that cluster variables can appear consecutively in the symbol for loop amplitudes only if there exists a cluster where they both appear, while for amplitudes at tree level it controls which poles can appear in individual BCFW terms. 
Until now, it was unclear whether cluster adjacency would also be present in the cosmological setting. In this paper we not only fill this gap but also show that cosmological correlators satisfy a stronger property, namely a generalisation of cluster adjacency, which we call the {\it ordered single cluster condition}: all letters appearing in a given symbol word must correspond to mutually compatible tubes, and their ordering reflects the inclusion relations among these tubes. This condition comes from the discontinuity structure of cosmological wavefunctions.  We show how powerful this condition is by bootstrapping the symbol of $P_{n\leq 4}$ and the star graph $S_4$, where the cluster constraints drastically reduce the space of allowed symbols. 

This paper is organized as follows. In section \ref{sec:recap} we will review all the notions needed for the derivations of differential equations and symbols for cosmological wavefunction coefficients. In section \ref{sec:alphabet}, we comment on the properties of the symbol alphabet for a given graph, and recall the definition of extended graphs. We highlight there the fact that each word in the symbol contains compatible tubes on the extended graph.
In section \ref{sec:Acluster} we state the relation between tubes on path graphs and edges and diagonals of an $n$-gon, showing that the compatibility on both sides matches. We conclude that each word of the symbol contains variables from a single cluster of $A_{2n-3}$  and provide an explicit embedding of alphabet letters into $\text{Gr}(2,2n)$.
In section \ref{sec:adj} we present a stronger version of cluster adjacency property. We explain that this cluster-like structure is also present for all graphs.
Finally, in section \ref{sec:bootstrap} we show how the symbol of wavefunction coefficients for tree graphs can be completely fixed using the integrability condition, the ordered single cluster condition and a limited number of basic physical constraints.

\paragraph{Note Added.}

While finishing this draft, the paper \cite{Paranjape:2026htn} appeared with similar results. 

\section{Review on Differential Equations and Symbols}\label{sec:recap}

In this paper we focus on the wavefunction coefficients in FRW cosmologies. Each such coefficient is associated to a graph $G$. To write explicit expressions, for a given graph $G$, we associate a variable $X_v$ to each vertex $v\in V_G$, and a variable $Y_e$ to each edge $e\in E_G$. Then the FRW wavefunction coefficient for a graph $G$ can be obtained by performing the twisted integral:
\begin{equation}\label{eq:FRWintro}
\psi_G^{\text{FRW}}(X_v,Y_e)=\int_0^\infty \left(\prod_v \dd x_v \,x_v^{\alpha_v}\right) \psi_G^{\text{flat}} (X_v+x_v,Y_e) \,,
\end{equation}
where the parameters $\alpha_v$ are related to the cosmological parameter $\epsilon$ and the type of interaction we consider (see \cite{Baumann:2024mvm} for more details). The flat-space functions $\psi_G^{\text{flat}} (X_v+x_v,Y_e)$ are rational functions of the kinematic variables associated with vertices and edges. In this paper, we focus our attention on de Sitter wavefunctions and $\phi^3$ interactions, in which case all $\alpha_v=0$, and the integrals reduce to
\begin{equation}\label{eq:dS1}
\psi_G^{\text{dS}}(X_v,Y_e)=\int_0^\infty \prod_v \dd x_v\, \psi_G^{\text{flat}} (X_v+x_v,Y_e)\,.
\end{equation}

As advocated in \cite{Capuano:2025ehm}, there exists a natural decomposition of the wavefunction $\psi_G^{\text{dS}}$ using functions labelled by tubings on the graph $G$. We recall here the definition of graph tubes and tubings following \cite{Glew:2025otn}. Given a graph $G$, a tube $T$ is the connected subgraph induced by a subset of its vertices $V_G$. The graph $G$ itself is also a tube which is called the \textit{root}. A $u$-tubing $\tau$ on $G$ is a collection of compatible tubes $\{T_i\}_{i=1,\ldots,p}$, where two tubes $T_i$ and $T_j$ are compatible if either one is included in another (i.e. $T_i\subset T_j$ or $T_j\subset T_i$), or they are not adjacent (i.e. the vertex set of $T_i\cup T_j$ does not induce a connected subgraph of $G$). A maximal set of compatible tubes is called a \textit{maximal} $u$-tubing. $\mathcal{U}_G^{\text{max}}$ is the set of all maximal $u$-tubings of $G$.

To each tube $T$ on $G$, we can associate the linear function
\begin{equation}\label{eq:HT}
H_T(X_v,Y_e)=\sum_{v\in V_T}\left(X_v+\sum_{e=[v,v'] \in G\setminus T} Y_e\right),
\end{equation}
where the second sum runs over the edges incident to $v$ in $G\setminus T$, and self loops are counted twice.
Moreover, for a given tubing $\tau$, we define a logarithmic form
\begin{equation}
    \mu_{\tau}(X_v,Y_e)=\bigwedge_{T\in\tau}\dlog{H_T}(X_v,Y_e)\,,
\end{equation}
where $\dd$ is defined to act only on the vertex variables $X_v$, and the order in the wedge product is fixed by an ordering of vertices of $G$ and a map $q_\tau$ that associates to $v\in V_G$ the smallest (by inclusion) tube $T_v\in\tau$ that contains $v$. Then, to each tubing we associate a function $F_\tau$ defined as
\begin{equation}
F_\tau(X_v,Y_e)=\int \mu_\tau(x_v+X_v,Y_e)
\end{equation}
where $\mu_\tau$ is now a form in the integration variables $\{x_v\}$, and the domain of integration is $\{x_v: 0< x_v<\infty\}$.

Using the functions $F_\tau$, the wavefunction coefficient $\psi_G^{\text{dS}}$ can be decomposed as the sum over subgraphs of $G$ obtained from it by cutting all possible subsets of its edges:
\begin{equation}\label{eq:decomp_psi}
    \psi_G^{\text{dS}}(X_v,Y_e)=\sum_{\mathbf{e}}(-1)^{|\mathbf{e}|}\sum_{\tau\in \mathcal{U}_{G_{\mathbf{e}}}^{\text{max}}}F_\tau\,,    
\end{equation}
where $G_\mathbf{e}=G[E_G\setminus \mathbf{e}]$.
For example, the wavefunction decomposition for the 2-site path graph reads
\begin{equation}
\psi_{\begin{tikzpicture}[scale=0.6]
    \coordinate (A) at (0,0);
    \coordinate (B) at (1,0);
    \draw[thick] (A) -- (B);
    \fill[black] (A) circle (2pt);
    \fill[black] (B) circle (2pt);
\end{tikzpicture}}^{\text{dS}}=F_{\begin{tikzpicture}[scale=0.6]
    \coordinate (A1) at (0,0);
    \coordinate (B1) at (1,0);
    \draw[thick] (A1) -- (B1);
        \draw[thick, black] (1/2,0) ellipse (0.75cm and 0.3cm);
            \draw[thick, black] (0,0) ellipse (0.15cm and 0.15cm);
    \fill[black] (A1) circle (2pt);
    \fill[black] (B1) circle (2pt);
\end{tikzpicture}}+F_{\begin{tikzpicture}[scale=0.6]
    \coordinate (A1) at (0,0);
    \coordinate (B1) at (1,0);
    \draw[thick] (A1) -- (B1);
        \draw[thick, black] (1/2,0) ellipse (0.75cm and 0.3cm);
            \draw[thick, black] (1,0) ellipse (0.15cm and 0.15cm);
    \fill[black] (A1) circle (2pt);
    \fill[black] (B1) circle (2pt);
\end{tikzpicture}}-F_{\begin{tikzpicture}[scale=0.6]
    \coordinate (A1) at (0,0);
    \coordinate (B1) at (1,0);
    \draw[thick] (A1) -- (B1);
            \draw[thick, black] (0,0) ellipse (0.15cm and 0.15cm);
            \draw[thick, black] (1,0) ellipse (0.15cm and 0.15cm);
    \fill[black] (A1) circle (2pt);
    \fill[black] (B1) circle (2pt);
\end{tikzpicture}}
\end{equation}

Using a variety of methods \cite{Arkani-Hamed:2023kig,Capuano:2025ehm}, it is possible to derive differential equations for the functions $F_\tau$ by acting with the kinematic differential $\dd_{\text{kin}}=\sum \frac{\partial}{\partial X_v}\dd X_v+\frac{\partial}{\partial Y_e}\dd Y_e$. The general form of these equations is
\begin{equation}
\dd_{\text{kin}}F_\tau=\sum_{\tau'}\dlog f_{\tau\tau'}F_{\tau'}\,,
\end{equation}
where the sum is over all possible $u$-tubings, not necessarily maximal, on $G_{\mathbf{e}}$ for all subsets of edges $\mathbf{e}\subseteq E_G$. The functions $ f_{\tau\tau'}$ have been identified in \cite{Capuano:2025myy} as the region variables of the graph $G$. We will not provide an explicit definition of regions in this paper, and we refer the interested reader to \cite{Capuano:2025ehm} instead.

For example, for the 2-site path graph, the set of all $u$-tubings, and therefore functions $F_\tau$, is
\begin{equation}
\vec F=\{F_{\begin{tikzpicture}[scale=0.6]
    \coordinate (A1) at (0,0);
    \coordinate (B1) at (1,0);
    \draw[thick] (A1) -- (B1);
        \draw[thick, black] (1/2,0) ellipse (0.75cm and 0.3cm);
            \draw[thick, black] (0,0) ellipse (0.15cm and 0.15cm);
    \fill[black] (A1) circle (2pt);
    \fill[black] (B1) circle (2pt);
\end{tikzpicture}},
F_{\begin{tikzpicture}[scale=0.6]
    \coordinate (A1) at (0,0);
    \coordinate (B1) at (1,0);
    \draw[thick] (A1) -- (B1);
        \draw[thick, black] (1/2,0) ellipse (0.75cm and 0.3cm);
            \draw[thick, black] (1,0) ellipse (0.15cm and 0.15cm);
    \fill[black] (A1) circle (2pt);
    \fill[black] (B1) circle (2pt);
\end{tikzpicture}},
F_{\begin{tikzpicture}[scale=0.6]
    \coordinate (A1) at (0,0);
    \coordinate (B1) at (1,0);
    \draw[thick] (A1) -- (B1);
            \draw[thick, black] (0,0) ellipse (0.15cm and 0.15cm);
            \draw[thick, black] (1,0) ellipse (0.15cm and 0.15cm);
    \fill[black] (A1) circle (2pt);
    \fill[black] (B1) circle (2pt);
\end{tikzpicture}},
F_{\begin{tikzpicture}[scale=0.6]
    \coordinate (A2) at (0,0);
    \coordinate (B2) at (1,0);
    \draw[thick] (A2) -- (B2);
    \draw[thick, black] (0,0) ellipse (0.15cm and 0.15cm);
    \fill[black] (A2) circle (2pt);
    \fill[black] (B2) circle (2pt);
\end{tikzpicture}},
F_{\begin{tikzpicture}[scale=0.6]
    \coordinate (A3) at (4,0);
    \coordinate (B3) at (5,0);
    \draw[thick] (A3) -- (B3);
    \draw[thick, black] (5,0) ellipse (0.15cm and 0.15cm);
    \fill[black] (A3) circle (2pt);
    \fill[black] (B3) circle (2pt);
\end{tikzpicture}},
F_{\begin{tikzpicture}[scale=0.6]
    \coordinate (A1) at (0,0);
    \coordinate (B1) at (1,0);
    \draw[thick] (A1) -- (B1);
        \draw[thick, black] (1/2,0) ellipse (0.75cm and 0.3cm);
    \fill[black] (A1) circle (2pt);
    \fill[black] (B1) circle (2pt);
    \end{tikzpicture}},
F_{\begin{tikzpicture}[scale=0.6]
    \coordinate (A) at (0,0);
    \coordinate (B) at (1,0);
    \draw[thick] (A) -- (B);
    \fill[black] (A) circle (2pt);
    \fill[black] (B) circle (2pt);
\end{tikzpicture}}\}
\end{equation}
where the last tubing is the empty tubing, and we set $F_{\begin{tikzpicture}[scale=0.6]
    \coordinate (A) at (0,0);
    \coordinate (B) at (1,0);
    \draw[thick] (A) -- (B);
    \fill[black] (A) circle (2pt);
    \fill[black] (B) circle (2pt);
\end{tikzpicture}}=1$.
In this example, the differential equations that we get are
\small
\begin{equation}
\dd_{\text{kin}}\left(\begin{matrix}
F_{\begin{tikzpicture}[scale=0.6]
    \coordinate (A1) at (0,0);
    \coordinate (B1) at (1,0);
    \draw[thick] (A1) -- (B1);
        \draw[thick, black] (1/2,0) ellipse (0.75cm and 0.3cm);
            \draw[thick, black] (0,0) ellipse (0.15cm and 0.15cm);
    \fill[black] (A1) circle (2pt);
    \fill[black] (B1) circle (2pt);
\end{tikzpicture}}\\
F_{\begin{tikzpicture}[scale=0.6]
    \coordinate (A1) at (0,0);
    \coordinate (B1) at (1,0);
    \draw[thick] (A1) -- (B1);
        \draw[thick, black] (1/2,0) ellipse (0.75cm and 0.3cm);
            \draw[thick, black] (1,0) ellipse (0.15cm and 0.15cm);
    \fill[black] (A1) circle (2pt);
    \fill[black] (B1) circle (2pt);
\end{tikzpicture}}\\
F_{\begin{tikzpicture}[scale=0.6]
    \coordinate (A1) at (0,0);
    \coordinate (B1) at (1,0);
    \draw[thick] (A1) -- (B1);
            \draw[thick, black] (0,0) ellipse (0.15cm and 0.15cm);
            \draw[thick, black] (1,0) ellipse (0.15cm and 0.15cm);
    \fill[black] (A1) circle (2pt);
    \fill[black] (B1) circle (2pt);
\end{tikzpicture}}\\
F_{\begin{tikzpicture}[scale=0.6]
    \coordinate (A2) at (0,0);
    \coordinate (B2) at (1,0);
    \draw[thick] (A2) -- (B2);
    \draw[thick, black] (0,0) ellipse (0.15cm and 0.15cm);
    \fill[black] (A2) circle (2pt);
    \fill[black] (B2) circle (2pt);
\end{tikzpicture}}\\
F_{\begin{tikzpicture}[scale=0.6]
    \coordinate (A3) at (4,0);
    \coordinate (B3) at (5,0);
    \draw[thick] (A3) -- (B3);
    \draw[thick, black] (5,0) ellipse (0.15cm and 0.15cm);
    \fill[black] (A3) circle (2pt);
    \fill[black] (B3) circle (2pt);
\end{tikzpicture}}\\
F_{\begin{tikzpicture}[scale=0.6]
    \coordinate (A1) at (0,0);
    \coordinate (B1) at (1,0);
    \draw[thick] (A1) -- (B1);
        \draw[thick, black] (1/2,0) ellipse (0.75cm and 0.3cm);
    \fill[black] (A1) circle (2pt);
    \fill[black] (B1) circle (2pt);
    \end{tikzpicture}}\\
F_{\begin{tikzpicture}[scale=0.6]
    \coordinate (A) at (0,0);
    \coordinate (B) at (1,0);
    \draw[thick] (A) -- (B);
    \fill[black] (A) circle (2pt);
    \fill[black] (B) circle (2pt);
\end{tikzpicture}}
\end{matrix}\right)=\left(\begin{matrix}
0&0&0&\dlog(X_2-Y_{1,2}) &0&\dlog\frac{(X_1+Y_{1,2})}{(X_2-Y_{1,2})}&0\\
0&0&0&0&\dlog(X_1-Y_{1,2})&\dlog\frac{(X_2+Y_{1,2})}{(X_1-Y_{1,2})}&0\\
0&0&0&\dlog(X_2+Y_{1,2})&\dlog(X_1+Y_{1,2})&0&0\\
0&0&0&0&0&0&\dlog(X_1+Y_{1,2})\\
0&0&0&0&0&0&\dlog(X_2+Y_{1,2})\\
0&0&0&0&0&0&\dlog(X_1+X_2)\\
0&0&0&0&0&0&0
\end{matrix}\right)\left(\begin{matrix}
F_{\begin{tikzpicture}[scale=0.6]
    \coordinate (A1) at (0,0);
    \coordinate (B1) at (1,0);
    \draw[thick] (A1) -- (B1);
        \draw[thick, black] (1/2,0) ellipse (0.75cm and 0.3cm);
            \draw[thick, black] (0,0) ellipse (0.15cm and 0.15cm);
    \fill[black] (A1) circle (2pt);
    \fill[black] (B1) circle (2pt);
\end{tikzpicture}}\\
F_{\begin{tikzpicture}[scale=0.6]
    \coordinate (A1) at (0,0);
    \coordinate (B1) at (1,0);
    \draw[thick] (A1) -- (B1);
        \draw[thick, black] (1/2,0) ellipse (0.75cm and 0.3cm);
            \draw[thick, black] (1,0) ellipse (0.15cm and 0.15cm);
    \fill[black] (A1) circle (2pt);
    \fill[black] (B1) circle (2pt);
\end{tikzpicture}}\\
F_{\begin{tikzpicture}[scale=0.6]
    \coordinate (A1) at (0,0);
    \coordinate (B1) at (1,0);
    \draw[thick] (A1) -- (B1);
            \draw[thick, black] (0,0) ellipse (0.15cm and 0.15cm);
            \draw[thick, black] (1,0) ellipse (0.15cm and 0.15cm);
    \fill[black] (A1) circle (2pt);
    \fill[black] (B1) circle (2pt);
\end{tikzpicture}}\\
F_{\begin{tikzpicture}[scale=0.6]
    \coordinate (A2) at (0,0);
    \coordinate (B2) at (1,0);
    \draw[thick] (A2) -- (B2);
    \draw[thick, black] (0,0) ellipse (0.15cm and 0.15cm);
    \fill[black] (A2) circle (2pt);
    \fill[black] (B2) circle (2pt);
\end{tikzpicture}}\\
F_{\begin{tikzpicture}[scale=0.6]
    \coordinate (A3) at (4,0);
    \coordinate (B3) at (5,0);
    \draw[thick] (A3) -- (B3);
    \draw[thick, black] (5,0) ellipse (0.15cm and 0.15cm);
    \fill[black] (A3) circle (2pt);
    \fill[black] (B3) circle (2pt);
\end{tikzpicture}}\\
F_{\begin{tikzpicture}[scale=0.6]
    \coordinate (A1) at (0,0);
    \coordinate (B1) at (1,0);
    \draw[thick] (A1) -- (B1);
        \draw[thick, black] (1/2,0) ellipse (0.75cm and 0.3cm);
    \fill[black] (A1) circle (2pt);
    \fill[black] (B1) circle (2pt);
    \end{tikzpicture}}\\
F_{\begin{tikzpicture}[scale=0.6]
    \coordinate (A) at (0,0);
    \coordinate (B) at (1,0);
    \draw[thick] (A) -- (B);
    \fill[black] (A) circle (2pt);
    \fill[black] (B) circle (2pt);
\end{tikzpicture}}
\end{matrix}\right)
\end{equation}
\normalsize

Knowing the explicit form of the differential equations, we immediately get a relation between symbols of the functions
\begin{equation}
\mathcal{S}(F_\tau)=\sum_{\tau'}\mathcal{S}(F_{\tau'})\otimes f_{\tau \tau'}\,.
\end{equation}
Importantly, since the resulting matrix of coefficients in the differential equations for the dS wavefunctions is strictly upper triangular, one can find the symbol for any $F_\tau$ by multiplying the matrix $|\tau|$ times. This implies that the symbol of any function $F_\tau$ has length $|\tau|$.

Combining the last observation with formula \eqref{eq:decomp_psi}, one can easily derive the symbol of  the dS wavefunction $\psi_{G}^{\text{dS}}$.

\section{Symbol Alphabet}
\label{sec:alphabet}
\begin{figure}[t]
\centering
\begin{tikzpicture}[scale=1.2]
    \coordinate (A1) at (0,0);
    \coordinate (B1) at (1,0);
    \coordinate (C1) at (2,0);
    \draw[thick] (A1) -- (B1)-- (C1);
    \draw (B1) node[line width=1pt] {$\times$};
    \fill[black] (A1) circle (2pt);
    \fill[black] (C1) circle (2pt);
    \node[] at (0,-0.23) {$1$};
    \node[] at(1,-0.23) {$(12)$};
    \node[] at (2,-0.23) {$2$};
\end{tikzpicture}
\caption{Extended $2$-site path graph.}
\label{fig:E2}
\end{figure}
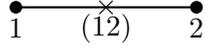

In the previous section we have reviewed a method to find the symbol of the de Sitter wavefunction coefficient $\psi_G^\text{dS}$. We observed that the entries of the symbol are region variables associated to regions on the graph $G$. Importantly, not all region variables are present in the final entries of the symbol, and only those corresponding to connected regions appear. This has already been observed in earlier works on the kinematic flow \cite{Arkani-Hamed:2023bsv,Baumann:2024mvm,Baumann:2025qjx}, and the authors of these papers introduced the notion of extended graph, which we now recall.
The extended graph $\widetilde{G}$ of the original graph $G$ can be obtained by adding a vertex in the middle of each edge of the graph $G$. 
We will indicate these new vertices with $\cross$ to distinguish them from the vertices of the graph $G$ that we denote with $\bullet$. For example, the extended graph of the $2$-site path graph is a $3$-site path graph depicted in Fig.~\ref{fig:E2}.
In general, the extended graph of graph $G$ will have $m=\abs{V_G}+\abs{E_G}$ vertices.

We now consider the tubes and tubings on the extended graph $\widetilde{G}$ since these tubes naturally encode all letters appearing in the symbol alphabet of cosmological wavefunctions.  More explicitly, to each tube $\widetilde{T}$ on $\widetilde{G}$, we associate a function
\begin{equation}
\widetilde{H}_{\widetilde{T}}=\sum_{i\in \widetilde{T}} \widetilde{H}_i,
\end{equation}
where for tubes containing just one vertex $\widetilde{H}_i\equiv \widetilde{H}_{\{i\}}$ we associate 
\begin{equation}
    \begin{cases}
    \widetilde{H}_{\bullet}=X_{\bullet}+\sum_{\cross \text{connected to }\bullet}Y_{\cross}\\
    \widetilde{H}_{\cross}=-2Y_{\cross}
    \end{cases}.
\end{equation}
With this definition, the root tube function $\widetilde{H}_{\widetilde{G}}=\sum_{i=1}^m \widetilde{H}_i$ represents the total energy. In particular, every tube $T$ on $G$ has a corresponding tube $\widetilde{T}$ on $\widetilde{G}$ such that $\widetilde{H}_{\widetilde{T}}=H_T$.

As a simple example,  on the extended graph in Fig.~\ref{fig:E2} of the 2-site path graph,  we can define $6$ tubes,  associated with $6$ variables
\begin{equation}
\begin{tabular}{lll}
\raisebox{-0.1cm}{\begin{tikzpicture}[scale=1.2]
    \coordinate (A1) at (0,0);
    \coordinate (B1) at (1,0);
    \coordinate (C1) at (2,0);
    \draw[thick] (A1) -- (B1)-- (C1);
    \draw[thick, black] (0,0) ellipse (0.15cm and 0.15cm);
    \draw (B1) node[line width=1pt] {$\times$};
    \fill[black] (A1) circle (2pt);
    \fill[black] (C1) circle (2pt);
\end{tikzpicture}}=$X_1+Y_{1,2}$,
&\raisebox{-0.1cm}{\begin{tikzpicture}[scale=1.2]
    \coordinate (A1) at (0,0);
    \coordinate (B1) at (1,0);
    \coordinate (C1) at (2,0);
    \draw[thick] (A1) -- (B1)-- (C1);
    
    \draw[thick, black] (B1) ellipse (0.15cm and 0.15cm);
  
    \draw (B1) node[line width=1pt] {$\times$};
    \fill[black] (A1) circle (2pt);
    \fill[black] (C1) circle (2pt);
\end{tikzpicture}}=$-2Y_{1,2}$,
&\raisebox{-0.1cm}{\begin{tikzpicture}[scale=1.2]
    \coordinate (A1) at (0,0);
    \coordinate (B1) at (1,0);
    \coordinate (C1) at (2,0);
    \draw[thick] (A1) -- (B1)-- (C1);

    \draw[thick, black] (C1) ellipse (0.15cm and 0.15cm);

    \draw (B1) node[line width=1pt] {$\times$};
    \fill[black] (A1) circle (2pt);
    \fill[black] (C1) circle (2pt);
\end{tikzpicture}}=$X_2+Y_{1,2}$\\
\hspace{-0.18cm}\raisebox{-0.2cm}{\begin{tikzpicture}[scale=1.2]
    \coordinate (A1) at (0,0);
    \coordinate (B1) at (1,0);
    \coordinate (C1) at (2,0);
    \draw[thick] (A1) -- (B1)-- (C1);
    \draw[thick, black] (1/2,0) ellipse (0.75cm and 0.25cm);

    \draw (B1) node[line width=1pt] {$\times$};
    \fill[black] (A1) circle (2pt);
    \fill[black] (C1) circle (2pt);
\end{tikzpicture}}=$X_1-Y_{1,2}$,
&\raisebox{-0.2cm}{\begin{tikzpicture}[scale=1.2]
    \coordinate (A1) at (0,0);
    \coordinate (B1) at (1,0);
    \coordinate (C1) at (2,0);
    \draw[thick] (A1) -- (B1)-- (C1);
    \draw[thick, black] (3/2,0) ellipse (0.75cm and 0.25cm);
    \draw (B1) node[line width=1pt] {$\times$};
    \fill[black] (A1) circle (2pt);
    \fill[black] (C1) circle (2pt);
\end{tikzpicture}}=$X_2-Y_{1,2}$,
&\hspace{-0.2cm}\raisebox{-0.2cm}{\begin{tikzpicture}[scale=1.2]
    \coordinate (A1) at (0,0);
    \coordinate (B1) at (1,0);
    \coordinate (C1) at (2,0);
    \draw[thick] (A1) -- (B1)-- (C1);
    \draw[thick, black] (B1) ellipse (1.25 cm and 0.25 cm);
    \draw (B1) node[line width=1pt] {$\times$};
    \fill[black] (A1) circle (2pt);
    \fill[black] (C1) circle (2pt);
\end{tikzpicture}}=$X_1+X_2$
\end{tabular}
\end{equation}
Then the symbol for the wavefunction coefficient for the 2-site path graph reads~\cite{Hillman:2019wgh}
\begin{equation}\label{eq:re2}
\begin{aligned}
\mathcal{S}(\psi^{\text{dS}}_{\begin{tikzpicture}[scale=0.6]
    \coordinate (A) at (0,0);
    \coordinate (B) at (1,0);
    \draw[thick] (A) -- (B);
    \fill[black] (A) circle (2pt);
    \fill[black] (B) circle (2pt);
\end{tikzpicture}})=&\frac{X_1+Y_{1,2}}{X_1+X_2}\otimes\frac{X_2-Y_{1,2}}{X_2+Y_{1,2}}+\frac{X_2+Y_{1,2}}{X_1+X_2}\otimes\frac{X_1-Y_{1,2}}{X_1+Y_{1,2}}\\
=&\frac{\hspace{-0.2cm}\begin{tikzpicture}[scale=1.2]
    \coordinate (A1) at (0,0);
    \coordinate (B1) at (1,0);
    \coordinate (C1) at (2,0);
    \draw[thick] (A1) -- (B1)-- (C1);
    \draw[thick, black] (0,0) ellipse (0.15cm and 0.15cm);
    \draw (B1) node[line width=1pt] {$\times$};
    \fill[black] (A1) circle (2pt);
    \fill[black] (C1) circle (2pt);
\end{tikzpicture}}{\begin{tikzpicture}[scale=1.2]
    \coordinate (A1) at (0,0);
    \coordinate (B1) at (1,0);
    \coordinate (C1) at (2,0);
    \draw[thick] (A1) -- (B1)-- (C1);
    \draw[thick, black] (B1) ellipse (1.25 cm and 0.25 cm);
    \draw (B1) node[line width=1pt] {$\times$};
    \fill[black] (A1) circle (2pt);
    \fill[black] (C1) circle (2pt);
\end{tikzpicture}}\otimes \frac{\begin{tikzpicture}[scale=1.2]
    \coordinate (A1) at (0,0);
    \coordinate (B1) at (1,0);
    \coordinate (C1) at (2,0);
    \draw[thick] (A1) -- (B1)-- (C1);
    \draw[thick, black] (3/2,0) ellipse (0.75cm and 0.25cm);
    \draw (B1) node[line width=1pt] {$\times$};
    \fill[black] (A1) circle (2pt);
    \fill[black] (C1) circle (2pt);
\end{tikzpicture}}{\hspace{-0.2cm}\begin{tikzpicture}[scale=1.2]
    \coordinate (A1) at (0,0);
    \coordinate (B1) at (1,0);
    \coordinate (C1) at (2,0);
    \draw[thick] (A1) -- (B1)-- (C1);

    \draw[thick, black] (C1) ellipse (0.15cm and 0.15cm);

    \draw (B1) node[line width=1pt] {$\times$};
    \fill[black] (A1) circle (2pt);
    \fill[black] (C1) circle (2pt);
\end{tikzpicture}}+\frac{\hspace{-0.2cm}\begin{tikzpicture}[scale=1.2]
    \coordinate (A1) at (0,0);
    \coordinate (B1) at (1,0);
    \coordinate (C1) at (2,0);
    \draw[thick] (A1) -- (B1)-- (C1);

    \draw[thick, black] (C1) ellipse (0.15cm and 0.15cm);

    \draw (B1) node[line width=1pt] {$\times$};
    \fill[black] (A1) circle (2pt);
    \fill[black] (C1) circle (2pt);
\end{tikzpicture}}{\begin{tikzpicture}[scale=1.2]
    \coordinate (A1) at (0,0);
    \coordinate (B1) at (1,0);
    \coordinate (C1) at (2,0);
    \draw[thick] (A1) -- (B1)-- (C1);
    \draw[thick, black] (B1) ellipse (1.25 cm and 0.25 cm);
    \draw (B1) node[line width=1pt] {$\times$};
    \fill[black] (A1) circle (2pt);
    \fill[black] (C1) circle (2pt);
\end{tikzpicture}}\otimes \frac{\begin{tikzpicture}[scale=1.2]
    \coordinate (A1) at (0,0);
    \coordinate (B1) at (1,0);
    \coordinate (C1) at (2,0);
    \draw[thick] (A1) -- (B1)-- (C1);
    \draw[thick, black] (1/2,0) ellipse (0.75cm and 0.25cm);

    \draw (B1) node[line width=1pt] {$\times$};
    \fill[black] (A1) circle (2pt);
    \fill[black] (C1) circle (2pt);
\end{tikzpicture}}{\hspace{-0.05cm}\begin{tikzpicture}[scale=1.2]
    \coordinate (A1) at (0,0);
    \coordinate (B1) at (1,0);
    \coordinate (C1) at (2,0);
    \draw[thick] (A1) -- (B1)-- (C1);
    \draw[thick, black] (0,0) ellipse (0.15cm and 0.15cm);
    \draw (B1) node[line width=1pt] {$\times$};
    \fill[black] (A1) circle (2pt);
    \fill[black] (C1) circle (2pt);
\end{tikzpicture}}\,,
\end{aligned}
\end{equation}
where we notice that the variable $-2Y_{1,2}$ never appears in the result.  

As for the original graph $G$, we can also study compatibility of tubes on the extended graph $\widetilde{G}$. The notion of compatibility that we will use in this case slightly differs from the one we introduced in section \ref{sec:recap}. We say that two tubes $\widetilde{T}_i$ and $\widetilde{T}_j$ on $\widetilde{G}$ are compatible\footnote{For interested readers, this notion of compatibility was introduced in \cite{Glew:2025otn} for the so-called $b$-tubes.} if either one is included in another (i.e. $\widetilde{T}_i\subset \widetilde{T}_j$ or $\widetilde{T}_j\subset \widetilde{T}_i$), or they are disjoint ($\widetilde{T}_i\cap \widetilde{T}_j=\emptyset$). Using this notion of compatibility, we state below the most important result of this paper:
\begin{center}
{\it The letters in each word in the symbol of $\psi_G^{\text{dS}}$ correspond to a set of compatible tubes on $\widetilde{G}$.}
\end{center}
This result is a significant generalisation of the usual statements in the literature on the symbol, which refer only to two neighbouring letters in symbol words. We have explicitly checked our generalised statement for symbols of various graphs, and we will provide the physical origin of this statement in section \ref{sec:adj}.

To support the above result, notice that for the 2-site path graph, there is only one pair of non-compatible tubes, namely
\begin{equation}
\left\{\begin{tikzpicture}[scale=1.2]
    \coordinate (A1) at (0,0);
    \coordinate (B1) at (1,0);
    \coordinate (C1) at (2,0);
    \draw[thick] (A1) -- (B1)-- (C1);
    \draw[thick, black] (1/2,0) ellipse (0.75cm and 0.25cm);

    \draw (B1) node[line width=1pt] {$\times$};
    \fill[black] (A1) circle (2pt);
    \fill[black] (C1) circle (2pt);
\end{tikzpicture},\begin{tikzpicture}[scale=1.2]
    \coordinate (A1) at (0,0);
    \coordinate (B1) at (1,0);
    \coordinate (C1) at (2,0);
    \draw[thick] (A1) -- (B1)-- (C1);
    \draw[thick, black] (3/2,0) ellipse (0.75cm and 0.25cm);
    \draw (B1) node[line width=1pt] {$\times$};
    \fill[black] (A1) circle (2pt);
    \fill[black] (C1) circle (2pt);
\end{tikzpicture}\right\}\,.
\end{equation}
The corresponding letters $X_1-Y_{1,2}$ and $X_2-Y_{1,2}$ indeed never appear in the same word in the symbol $\mathcal{S}(\psi^{\text{dS}}_{\begin{tikzpicture}[scale=0.6]
    \coordinate (A) at (0,0);
    \coordinate (B) at (1,0);
    \draw[thick] (A) -- (B);
    \fill[black] (A) circle (2pt);
    \fill[black] (B) circle (2pt);
\end{tikzpicture}})$ in \eqref{eq:re2}.  

\section{Relation to Type-$A$ Cluster Variables}
\label{sec:Acluster}
In this section we will focus our attention on path graphs, for which we previously established a relation between the  symbol alphabet and cluster algebras of type $A$. We will provide a modified version of our previous efforts, and show how one can associate a cluster algebra $A_{2n-3}$ with coefficients to a path graph $P_n$ with $n$ vertices. This differs from the results that we presented in \cite{Capuano:2025myy}, see also \cite{Mazloumi:2025pmx,Paranjape:2026htn}, where the associated algebra was $A_{2n-2}$.

For a path graph $G=P_n$ with $n$ vertices, the extended graph $\widetilde{G}=\widetilde{P}_n\cong P_{2n-1}$ is isomorphic to a path graph with $2n-1$ vertices. To simplify our combinatorial considerations, we will label this extended graph as
\begin{center}
\begin{tikzpicture}[scale=1]
    
    \draw[thick] (0,0) -- (2,0)-- (4,0)--(6,0);
    
    \node at (7,0) {$\cdots$};
    
    \draw[thick] (8,0) -- (10,0)-- (12,0)--(14,0);
  
    \fill[black] (0,0) circle (2pt);
    \draw (2,0) node[line width=1pt] {$\times$};
    \fill[black] (4,0) circle (2pt);
    \draw (6,0) node[line width=1pt] {$\times$};

    \draw (8,0) node[line width=1pt] {$\times$};
    \fill[black] (10,0) circle (2pt);
    \draw (12,0) node[line width=1pt] {$\times$};
     \fill[black] (14,0) circle (2pt);

    \node[scale=0.7] at (0,-0.3) {$1$};
    \node[scale=0.7] at(2,-0.3) {$2$};
    \node[scale=0.7] at (4,-0.3) {$3$};
    \node[scale=0.7] at(6,-0.3) {$4$};
    
    \node[scale=0.7] at (8,-0.3) {$2n-4$};
    \node[scale=0.7] at (10,-0.3) {$2n-3$};
    \node[scale=0.7] at (12,-0.3) {$2n-2$};
    \node[scale=0.7] at(14,-0.3) {$2n-1$};

\end{tikzpicture}
\end{center}
In this case, any tube on the extended graph can be labelled as $\widetilde{T}_{[i,j]}=\{i,i+1,\ldots,j-1,j\}$, and it is just an interval on the line. There is a natural bijection between all tubes on the extended graph $P_{2n-1}$ and edges and chords in a $(2n)$-gon, which we label $d_{ij}$, given by
\begin{equation}\label{eq:Ttod}
\widetilde{T}_{[i,j]}\equiv d_{i,j+1} \,.
\end{equation}
This bijection has the property that compatibility of tubes on $P_{2n-1}$ is equivalent to a non-crossing condition of the corresponding edges and chords. Therefore any collection of compatible tubes on $P_{2n-1}$ corresponds to a collection of non-intersecting chords, a partial triangulation of the $2n$-gon. Moreover, there is a natural relation between chords in an $m$-gon and the cluster algebra $A_{m-3}$, which we can use to map the symbol alphabet for $\psi_{P_n}^{\text{dS}}$ to the cluster structure of the cluster algebra $A_{2n-3}$.

To illustrate it, let us take the simplest example of the $2$-site path graph $P_2$. As we have already mentioned, there are 6 tubes on the extended graph $\widetilde{P}_2\cong P_3$, and they can be mapped to the 4 edges and 2 diagonals of a quadrilateral. Moreover, it is easy to check that the corresponding 6 tube variables satisfy the following quadratic relation
\begin{equation}
(X_1-Y_{1,2})(X_2-Y_{1,2})=(X_1+Y_{1,2})(X_2+Y_{1,2})+(-2Y_{1,2})(X_{1}+X_2)\,,
\end{equation}
which is precisely the exchange relation of the $A_1$ cluster algebra where $X_1-Y_{1,2}$ and $X_2-Y_{1,2}$ are the two mutable cluster variables of $A_1$, and the remaining 4 variables, associated with the singleton tubes and the root, are the 4 frozen cluster variables. The exchange relation can be illustrated on the quadrilateral as the following diagonal flip
\begin{equation}
\includegraphics[align=c,scale=1.2]{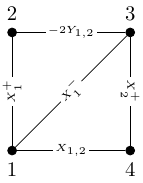}\qquad\leftrightarrow\qquad\includegraphics[align=c,scale=1.2]{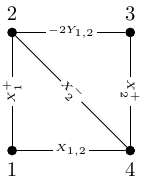}
\end{equation}

In general,  among the $(2n-1)n$ tube variables of the extended graph for $P_n$,  the $(2n-1)$ tubes containing just one vertex, together with the root, are the frozen variables, and they correspond to the edges of the $(2n)$-gon. The remaining variables correspond to $(2n-3)n$ chords in the $2n$-gon, which is exactly the number of mutable cluster variables of $A_{2n-3}$.  
 
\begin{figure}[H]
\centering
\includegraphics[]{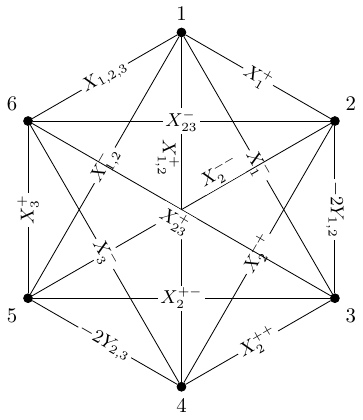}
\caption{Hexagon corresponding to extended $P_3$. }
\label{fig:hex}
\end{figure}

A further example is  given by the path graph with 3 vertices, for which the extended graph is isomorphic to $P_5$, and each tube on the extended graph can be mapped using \eqref{eq:Ttod} to the hexagon depicted in Fig.~\ref{fig:hex}.

\begin{figure}[H]
\centering
\includegraphics[scale=2.3]{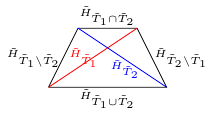}
\caption{General exchange relation for the tube functions on the extended graph.} 
\label{fig:Pto}
\end{figure}

As previously stated, under this correspondence compatibility of tubes is mapped precisely to non-crossing diagonals. 
Two tubes $\widetilde{T}_{[i,j]}$ and $\widetilde{T}_{[k,l]}$ are incompatible if and only if the diagonals $d_{i,j+1}$ and $d_{k,l+1}$ cross. On the algebraic level,  since $\widetilde{H}_{\widetilde{T}}=\sum_{i\in \widetilde{T}} \widetilde{H}_i$, any two variables corresponding to non-compatible tubes $\widetilde{T}_1,\widetilde{T}_2$ manifestly satisfy the following quadratic relation
\begin{equation}
\widetilde{H}_{\widetilde{T}_1} \widetilde{H}_{\widetilde{T}_2}=\widetilde{H}_{\widetilde{T}_1\cup \widetilde{T}_2}\widetilde{H}_{\widetilde{T}_1\cap \widetilde{T}_2}+\widetilde{H}_{\widetilde{T}_1\setminus \widetilde{T}_2}\widetilde{H}_{\widetilde{T}_2\setminus \widetilde{T}_1},
\end{equation}
which is exactly the Ptolemy's exchange relation of the type $A$ cluster algebra, as illustrated in Fig.~\ref{fig:Pto}.

Finally, to end this section, we embed the symbol alphabet, which coincides with the cluster variables of the $A_{2n-3}$ cluster algebra, into the Grassmannian $\text{Gr}(2,2n)$. This construction follows the standard realization of finite-type cluster algebras in terms of configuration spaces of points on $\mathbb{CP}^1$, as discussed for example in \cite{Arkani-Hamed:2020tuz}. In particular, the cluster algebra $A_{2n-3}$ is naturally associated with the moduli space $\mathcal{M}_{0,2n}$, the configuration space of $2n$ punctures on $\mathbb{CP}^1$ modulo projective transformations. Concretely, one may represent a point in $\mathcal{M}_{0,2n}$ by $2n$  coordinates $\{z_i\}_{i=1,2,\ldots,2n}$ defined up to a $\mathrm{SL}(2)$ transformation which allows us to gauge fix three $z_i$'s. In the Grassmannian description this corresponds to a $2\times 2n$ matrix representative of $\text{Gr}(2,2n)$ 
\begin{equation}
C=\left(\begin{matrix}
1&1&1&\ldots&1&1\\
z_1&z_2&z_3&\ldots&z_{2n-1}& z_{2n}
\end{matrix}\right),
\end{equation}
where the Pl\"ucker coordinates $(ij)$ of $\text{Gr}(2,2n)$ become the difference of puncture coordinates $(ij)=z_j-z_i$.
To make contact with the tube variables introduced earlier, we now give an explicit parametrization of the coordinates $z_i$ in terms of the tube variables. In this parametrization the Grassmannian representative takes the form
\begin{equation}
C=\left(\begin{matrix}
1&1&1&1&\ldots&1&1\\
0&X_1+Y_{1,2}&X_1-Y_{1,2}&X_1+X_2+Y_{2,3}&\ldots &X_{1}+\ldots+X_{n-1}-Y_{n-1,n}&X_1+\ldots+X_n
\end{matrix}\right),
\end{equation}
where we fix $z_1$ to be 0 and $z_2,z_{2n}$ to be two frozen variables $X_1+Y_{1,2}$ and $\sum_{i=1}^n X_i$, respectively.  Consequently, the Pl\"ucker coordinates $(ij)$ reproduce precisely the symbol alphabet of the path graph, providing a concrete embedding of the symbol alphabet into the Grassmannian $\text{Gr}(2,2n)$.

In the table below we summarize the three languages: tubes on extended graphs, chords in the polygon, and Grassmannian cluster structure.
\begin{table}[H]
  \centering
    \begin{tabular}{|c|c|c|c|}
    \hline
          & Symbol letters & Frozen variables & Compatibility \\\hline
    Extended Graph $P_{2n-1}$ & Tube $\widetilde{T}_{[i,j]}$ & Singlet tubes and root & compatible tubes \\\hline
    $2n$-gon &  chords $d_{i,j+1}$ & edges $e_{i,i+1}$ & non-intersecting diagonals \\\hline
    $\text{Gr}(2,2n)$ & Minor $(i,j+1)$ & $(i,i+1)$ & non-overlapping minors \\\hline
    \end{tabular}%
      \caption{Equivalent languages for symbol alphabet of $P_n$.}
  \label{tab:3pn}%
\end{table}%

\section{Generalized Cluster Adjacency and Its Origin}\label{sec:adj}
In this section, we will first provide a physical origin for the generalised cluster adjacency we stated in the previous sections. Additionally, we will state a stronger version of this condition, which we call the \emph{\textbf{ordered single cluster condition}}.  This condition will play a crucial role in the symbol bootstrap discussed in the next section.  

As in the SYM case, the cluster adjacency of cosmological wavefunctions is closely related to extended Steinmann relations~\cite{Caron-Huot:2019bsq}. The original Steinmann relation~\cite{Steinmann:1960soa,Steinmann:1960sob} states that the double discontinuity of two incompatible Mandelstams must vanish. While the Steinmann relations were originally proposed as constraints on the first two discontinuities of amplitudes in SYM theory, they have been extended to all depths in the symbol of amplitudes and form factors~\cite{Caron-Huot:2020bkp,Bossinger:2025rhf} and Feynman integrals~\cite{Chicherin:2020umh,He:2021mme}. 

For cosmological wavefunctions, discontinuity structures have been studied in~\cite{Hillman:2019wgh,Benincasa:2020aoj}.  In particular, the discontinuities of the wavefunction occur when one of the tube variables of the graph $G$ becomes negative. For a given tube $T$ on $G$, the corresponding function is always of the form \eqref{eq:HT}. Importantly, there exists a tube $\widetilde{T}$ on the extended graph $\widetilde{G}$ that has the same tube function $\widetilde{H}_{\widetilde{T}}=H_T$.  Diagrammatically, the discontinuity around  $H_{T}=0$ splits the graph into connected components as follows
\begin{figure}[H]
    \centering
    \includegraphics[width=0.5\linewidth]{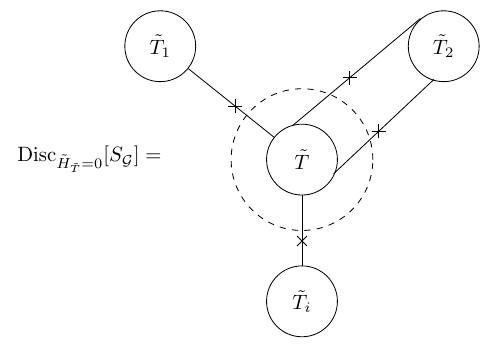}
    \caption{Discontinuity of the cosmological wavefunction.}
    \label{fig:disc}
\end{figure}
Explicitly, at the level of symbols, we have \cite{Hillman:2019wgh}
\begin{equation}\label{eq:disc}
\begin{aligned}
\underset{H_{T}=0}{\mathrm{Disc}}\!\left[\mathcal{S}(\psi_G^\text{dS}(X_v,Y_e))\right]=
\underset{H_T=0}{\mathrm{Disc}}\!\left[\mathcal{S}(\psi_T^{\text{dS}}(X'_v,Y_e))\right]\shuffle_i \sum_{\{\sigma_{e_c}\}}
(-1)^{n_+}\,
\mathcal{S}\!\left(\psi_{T_i}^{\text{dS}}(X''_v,Y_e)\right)
\end{aligned}
\end{equation}
where 
\begin{equation}
X'_v=\begin{cases}X_{v_c}+Y_{e_c},&\text{if the edge $e_c$ is cut by $T$, and $v_c$ is an endpoint of $e_c$ }\\X_v,&\text{otherwise}\end{cases}
\end{equation}
and 
\begin{equation}
X''_v=\begin{cases}X_{v_c}+\sigma_{e_c}Y_{e_c},&\text{if the edge $e_c$ is cut by $T$, and $v_c$ is an endpoint of $e_c$ }\\X_v,&\text{otherwise}\end{cases}
\end{equation}
Here, $X_{v_c}$ belongs to the set of vertex energies in a given subgraph corresponding to vertices with cut edges ending on them, and the $Y_{e_c}$ are the corresponding edge energies. The $\sigma_{e_c}$ can take on values $\pm1$ and the sum is over all such combinations, and $n_+$ counts the number of $+1$'s among the $\sigma_{e_c}$.

First of all, the discontinuity condition implies that the first entry must be of the form $H_T$ as in \eqref{eq:HT}. After taking this discontinuity, which removes the first entry of the symbol, we can consider further discontinuities around $\widetilde{H}_{\widetilde{T}'}=0$, where now some terms with $-Y$ can appear in the expressions for $\widetilde{H}_{\widetilde{T}'}$. Taking such discontinuities again leads to a shuffle product among the connected components similar to~\eqref{eq:disc}.

From Fig.~\ref{fig:disc} one can see that after taking the discontinuity around $H_T=0$, the symbol factorizes into a shuffle product of terms associated with the individual connected components. This means that if we want to take further discontinuities, they will be located at $\widetilde{H}_{\widetilde{T}'}=0$ for some tube $\widetilde{T}'$ that must be compatible with the tube $\widetilde{T}$ as tubes on extended graphs.

In fact, Fig.~\ref{fig:disc} reveals a stronger adjacency constraint between consecutive symbol letters. Suppose a symbol contains $\ldots\otimes \widetilde{H}_{\widetilde{T}_1}\otimes\ldots \otimes \widetilde{H}_{\widetilde{T}_2}\otimes\ldots$, which corresponds to taking the discontinuity $\widetilde{H}_{\widetilde{T}_1}=0$ before the discontinuity $\widetilde{H}_{\widetilde{T}_2}=0$. By the argument above, this is only possible if $\widetilde{T}_2$ is compatible with $\widetilde{T}_1$. Therefore, all letters appearing within a symbol word must correspond to mutually compatible tubes. We emphasize that this property does not hold for symbols of amplitudes or Feynman integrals of other theories such as SYM~\cite{Dixon:2016nkn,Chicherin:2020umh,He:2021mme}, ABJM~\cite{Caron-Huot:2012sos}, or QCD~\cite{Carrolo:2026qpu}.

Moreover, for two compatible tubes with $\widetilde{T}_1 \supset \widetilde{T}_2$, suppose we have already taken the discontinuity around the channel $\widetilde{H}_{\widetilde{T}_1}=0$. From \eqref{eq:disc}, we see that the function $\widetilde{H}_{\widetilde{T}_2}$, corresponding to a subgraph of $\widetilde{T}_1$, can only appear in the first factor of the equation \eqref{eq:disc}.  This implies that $\widetilde{H}_{\widetilde{T}_2}$ can appear in the symbol {\it after} $\widetilde{H}_{\widetilde{T}_1}$, but not before it. These two conditions together form what we call the \emph{\textbf{ordered single cluster condition}}.

\section{Cluster Bootstrap}\label{sec:bootstrap}
Having established the cluster conditions in the previous section, we will now supplement them with additional basic conditions that the symbol for $\psi_G^{\text{dS}}$ needs to satisfy, which will in turn allow us to bootstrap the symbol without referring back to differential equations or recursion relations. We start our bootstrap of the symbol by writing down an ansatz
\begin{equation}
    \mathcal{S}(\psi_{G}^\text{dS})=\sum_{\alpha_1,\ldots,\alpha_n} C_{\alpha_1,\dots,\alpha_n}\phi_{\alpha_1}\otimes\ldots\otimes\phi_{\alpha_n}\,,
\end{equation}
where $\phi_{\alpha_i}$'s belong to the symbol alphabet of $G$, and $C_{\alpha_1,\ldots,\alpha_n}$ are some unknown coefficients. The main question is then how to constrain the coefficients $C_{\alpha_1,\ldots,\alpha_n}$ using properties of the symbol and the cluster conditions discussed above.

We start fixing the coefficients of the ansatz by considering the first entry.  As a consequence of locality, the wavefunction can only have singularities when the total energy of a connected subgraph vanishes, 
{\it i.e.}
$H_T=\sum_{v\in T}X_v+\sum_{e\in e_c}Y_{e}=0$. This constrains the first entry of the symbol, leading to the so-called {\it first entry condition} 
\begin{equation}
    \text{First entry of } \mathcal{S}(\psi_G^{\text{dS}})\in\{H_T=\sum_{v\in T}X_v+\sum_{e\in e_c}Y_{e}: T\text{ is a tube on }G\}\,.
\end{equation}
For path graphs,  the first entry is therefore of the form $\sum_{k=i}^j X_k+Y_{i-1,i}+Y_{j,j+1}$.

The second condition that we impose on the symbol is the {\it integrability condition}: the function must have commuting second derivatives with respect to any two independent variables $x_i$ and $x_j$
\begin{equation}
    \frac{\partial^2 F}{\partial x_i\partial x_j}=\frac{\partial^2 F}{\partial x_j\partial x_i}\,.
\end{equation}
Integrability is equivalent to the following condition on each pair of adjacent slots:
\begin{equation}
\begin{aligned}
    &\sum_{\alpha_1,\ldots,\alpha_n} C_{\alpha_1,\dots,\alpha_n}\phi_{\alpha_1}\otimes\ldots\otimes\phi_{\alpha_n}\text{ is integrable}\\
    \Leftrightarrow 
    &\sum_{\alpha_1,\dots,\alpha_n}
C_{\alpha_1,\dots,\alpha_n}\,
\underbrace{\phi_{\alpha_1}\otimes\dots\otimes \phi_{\alpha_n}}_{\text{omitting }\alpha_j,\alpha_{j+1}}
\, \dlog \phi_{\alpha_j}\wedge \dlog \phi_{\alpha_{j+1}}
=0 \qquad \forall j\in\{1,2,\dots,n-1\}\,.
\end{aligned}
\end{equation}
After expanding all possible $\dlog\phi_{i}\wedge\dlog\phi_j$ into some basis, this leads to a set of linear equations for the coefficients.  Remarkably, we note that such conditions have a natural interpretation from the tubing picture, which also gives an invariant formulation without referring to any explicit parametrization:  a complete set of independent relations among $\dlog$
two-forms is generated from the tube pictures by the relations
\begin{equation}
\dlog \widetilde{H}_{{\widetilde{T}}_1}\wedge \dlog \widetilde{H}_{\widetilde{T}_1\cup\widetilde{T}_2}=\dlog \widetilde{H}_{\widetilde{T}_2}\wedge \dlog \widetilde{H}_{\widetilde{T}_1\cup\widetilde{T}_2}+\dlog \widetilde{H}_{{\widetilde{T}}_1}\wedge \dlog \widetilde{H}_{{\widetilde{T}_2}}
\end{equation}
for any pair of disjoint tubes $\widetilde{T}_1$ and $\widetilde{T}_2$. 
For details on how to construct integrable symbols, see~\cite{Dixon:2016nkn,He:2025tyv}.

\begin{figure}
\begin{center}\includegraphics[scale=1]{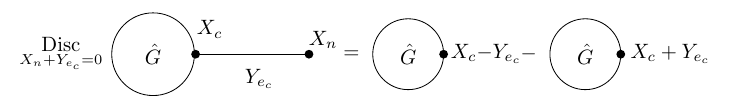}\end{center}
\caption{Discontinuity condition for symbols.}
\label{fig:disc2}
\end{figure}

We supplement the first-entry and integrability conditions with the cluster condition we studied in the previous section. Moreover, we impose a discontinuity condition which states that
\begin{equation}\label{eq:discn}
    \underset{X_n+Y_{e_c}=0}{\mathrm{Disc}} \mathcal{S}(\psi_G^\text{dS})=\left. \mathcal{S}(\psi_{\hat{G}}^{\text{dS}})\right|_{X_{c}\to X_{c}-Y_{e_c}} - \left. \mathcal{S}(\psi_{\hat{G}}^{\text{dS}})\right|_{X_{c}\to X_{c}+Y_{e_c}}
\end{equation}
where the vertex $n$ is a leaf of graph $G$ and the graph $\hat{G}$ is obtained from the graph $G$ by removing vertex $n$. Fig.~\ref{fig:disc2} provides a pictorial representation of this relation.
We claim that these four conditions completely fix the symbol of $\psi_G^{\text{dS}}$, at least for graphs with a small number of vertices.

Let us illustrate our approach for the 2-site path graph as a warm-up example. In this case, the first entry condition indicates that the only letters that can appear in the first entry of $\mathcal{S}(\psi_{P_2}^{\text{dS}})$ are $X_1+Y_{1,2}$, $X_2+Y_{1,2}$ or $X_1+X_2$. Together with the integrability condition, it leaves a five-dimensional space of integrable weight-2 symbols, and our ansatz becomes
\begin{equation}\label{eq:ansatz5}
\begin{aligned}
\mathcal{S}(\psi_{P_2}^{\text{dS}})=&c_5 \frac{X_1+Y_{1,2}}{X_1+X_2}\otimes\frac{X_2-Y_{1,2}}{X_1+X_2}+c_1 \frac{X_1+Y_{1,2}}{X_1+X_2}\otimes\frac{X_1+Y_{1,2}}{X_1+X_2}+c_3 \frac{X_2+Y_{1,2}}{X_1+X_2}\otimes\frac{X_1-Y_{1,2}}{X_1+X_2}\\
+&c_2 \left(\frac{X_1+Y_{1,2}}{X_1+X_2}\otimes\frac{X_2+Y_{1,2}}{X_1+X_2}+\frac{X_2+Y_{1,2}}{X_1+X_2}\otimes\frac{X_1+Y_{1,2}}{X_1+X_2}\right)+c_4 \frac{X_2+Y_{1,2}}{X_1+X_2}\otimes\frac{X_2+Y_{1,2}}{X_1+X_2} \,.
\end{aligned}
\end{equation}

Starting from this combination, and imposing the ordered single cluster condition, we fix all but one coefficient in \eqref{eq:ansatz5}
\begin{align}
 \mathcal{S}(\psi_{P_2}^{\text{dS}})=   c_2 \left(-\frac{X_1+Y_{1,2}}{X_1+X_2}\otimes\frac{X_2-Y_{1,2}}{X_1+X_2}+\frac{X_1+Y_{1,2}}{X_1+X_2}\otimes\frac{X_2+Y_{1,2}}{X_1+X_2}\right. \nonumber\\ \left.
 -\frac{X_2+Y_{1,2}}{X_1+X_2}\otimes\frac{X_1-Y_{1,2}}{X_1+X_2}+\frac{X_2+Y_{1,2}}{X_1+X_2}\otimes\frac{X_1+Y_{1,2}}{X_1+X_2}\right) \,.
\end{align}
To fix the overall coefficient, we can use the discontinuity condition \eqref{eq:discn} and we find $c_2=-1$.

This bootstrap procedure works efficiently for any tree graph. In particular, the ordered single cluster condition reduces the number of coefficients significantly in all cases we studied. We summarise in Table~\ref{tab:countboots} the reduction we encounter for the graphs we studied. For example, the 4-site star graph $S_4$ has an alphabet consisting of 33 letters. This alphabet is a subset of the 36 tubes on the extended star graph, obtained by removing the three tubes corresponding to 
$-2Y_{i,j}$, which never appear.
\begin{table}[H]
  \centering
    \begin{tabular}{|c|c|c|c|}\hline
    Graph    & Int+First&Cluster compatibility&  Discontinuity \\\hline
    $P_2$     & 5         & 1         & 0 \\\hline
    $P_3$   & 98        & 3        & 0 \\\hline
    $P_4$     & 2536     &8       & 0 \\\hline
     $S_4$     & 3522     &7       & 0 \\\hline
    \end{tabular}%
    \caption{The dimensions of the ansatz space after imposing constraints.}
  \label{tab:countboots}%
\end{table}%

As shown in Table~\ref{tab:countboots}, the adjacency constraint turns out to be extremely restrictive.  Starting from the integrability and first-entry conditions, the dimension of the ansatz space grows rapidly with the size of the graph. For instance, for the path graph $P_4$ the space contains 2536 possible integrable symbols. However, once the ordered single cluster condition is imposed, the dimension collapses to only 8. A similar drastic reduction occurs for the star graph $S_4$, where the ansatz space shrinks from 3522 to 7. In both cases, the ordered single cluster condition removes about $99.7\%\sim 99.8\%$ of the degrees of freedom.  We also comment here that this condition already implies the final entry condition: the final entry can only be a tube variable involving a single vertex of $G$  {\it i.e.} $X_i+\sum_{e_c}\sum_{\sigma_{e_c}}\sigma_{e_c}Y_{e_c}$.  This condition can also be read off from differential equations~\cite{Arkani-Hamed:2023kig,Arkani-Hamed:2023bsv}. 

\section{Summary and Outlook}
In this work we investigated the cluster algebra structure underlying cosmological wavefunction coefficients in conformally coupled scalar theories in de Sitter space. We showed that the symbol alphabet of these wavefunctions admits a natural combinatorial description in terms of tubes on the extended graph. For path graphs $P_n$ this structure is equivalent to the cluster algebra $A_{2n-3}$, where compatible tubes correspond to compatible cluster variables. Furthermore, we proposed the ordered single cluster condition, i.e. a generalized version of cluster adjacency, which means all letters appearing in a single word correspond to mutually compatible tubes, and their ordering is related to the inclusion structure of these tubes.  Together with integrability, first entry and discontinuity conditions, we fixed the symbol for graphs up to the $4$-site path and the $4$-site star. Remarkably, the ordered single cluster condition is extremely restrictive: it reduces the dimension of the space of integrable symbols to only about  $0.2\sim 0.3\%$.

From a mathematical perspective, an important open question is to understand cluster-algebra-like structures for more general graphs. Combinatorially, the symbol alphabet of the wavefunction is closely related to graph associahedra, whose boundary structures have been studied both from a combinatorial point of view~\cite{postnikov2005permutohedraassociahedra,Manneville_2017} as well as in terms of $u$-equations and binary geometries~\cite{Arkani-Hamed:2019mrd,He:2020onr}. It would be interesting to explore these connections further and we leave this for future research.

Once the symbol for general graphs is understood, another natural direction is to investigate patterns in the coefficients in front of each word in the symbol. Such patterns may reveal deeper structures, including more general properties that could be further explored by machine learning~\cite{Cai:2024znx,Basso:2024hlx}.

Finally, it would be interesting to extend these ideas beyond tree level. Applying similar bootstrap techniques to loop diagrams may reveal new analytic structures.

\section{Acknowledgements}

This work was supported in part by the Deutsche Forschungsgemeinschaft (DFG, German Research Foundation) Projektnummer 508889767/FOR5582
``Modern Foundations of Scattering Amplitudes''.

\bibliographystyle{nb}

\bibliography{cosmo}

\end{document}